\documentclass[preprint,authoryear,12pt]{elsarticle}
\usepackage{natbib}
\usepackage{amssymb}
\usepackage{color}
\begin{document}
\bibliographystyle{plainnat}
\begin{frontmatter}
\title{Unified Description of Matrix Mechanics and Wave Mechanics II}
\author{Yongqin Wang$^{1}$, Lifeng Kang$^{2}$}
\address{$^{1}$Department of Physics, Nanjing University, Nanjing 210008, China\\e-mail:yhwnju@hotmail.com}
\address{$^{2}$School of Pharmacy, Faculty of Medicine and Health, University of Sydney, NSW 2006, Australia}
\end{frontmatter}
\emph{Continue}\\
\setcounter{section}{4}
\setcounter{equation}{37}
\section{\bf The Hydrogen Atom}
\label{sec:1}
The Hamiltonian of the hydrogen atom can be written as
\begin{displaymath}
\hat{H}=\frac{\hat{p}^{2}}{2\mu}-\frac{k}{r}
\end{displaymath}
where $\mu$ is the reduced mass of the positive proton and negative electron, k is a parameter. There is
\begin{displaymath}
[\hat{L}_{z},\hat{H}]=[\hat{L}^{2},\hat{H}]=[\hat{L}_{z},\hat{L}^{2}]=0
\end{displaymath}
Hence, the operators $\hat{H}, \hat{L}_{z}$ and $\hat{L}^2$ have the orthonormalized simultaneous eigenfunctions $\psi$. In the spherical coordinate, the Hamiltonian operator of hydrogen atom is written as
\begin{displaymath}
\hat{H}=-\frac{\hbar^2}{2\mu}(\frac{\partial^{2}}{\partial r^{2}}+\frac{2}{r}\frac{\partial}{\partial r})+\frac{\hat{L}^{2}}{2\mu r^{2}}-\frac{k}{r}
\end{displaymath}
Using (32) and (37), we have for the Schr$\ddot{o}$dinger equation
\begin{equation}
[-\frac{\hbar^2}{2\mu}(\frac{\partial^{2}}{\partial r^{2}}+\frac{2}{r}\frac{\partial}{\partial r})+\frac{l(l+1)\hbar^{2}}{2\mu r^{2}}-\frac{k}{r}]\psi
=\hat{H}\psi=E\psi
\end{equation}
The eigenfunctions $\psi$ are capable of the separation of variables. Therefore
\begin{equation}
\psi=RY_{lm}
\end{equation}
Using (39) in (38), we then divide both sides by $Y_{lm}$ to get an ordinary differential equation for the unknown function $R=R(r)$
\begin{equation}
-\frac{\hbar^2}{2\mu}(\frac{d^{2}R}{dr^{2}}+\frac{2}{r}\frac{dR}{dr})+\frac{l(l+1)\hbar^{2}}{2\mu r^{2}}R-\frac{k}{r}R=ER
\end{equation}
Scheme I\\
The second order differential equation (40) is solved and we will find (86). So (39) becomes
\begin{small}
	\begin{equation}
	\begin{array}{cc}
	\psi_{nlm}=R_{nl}Y_{lm} & (n=1,2,\cdots;l=0,1,\cdots,n-1;m=-l,1-l,\cdots,l)
	\end{array}
	\end{equation}
\end{small}
where the radial wave functions $R_{nl}$ and spherical harmonic functions $Y_{lm}$ are respectively given by (36) and (93).\\
The Runge-Lenz vector operator of the hydrogen atom $^{[4]}$ is defined by
\begin{displaymath}
\hat{\vec{A}}=\frac{1}{\mu}(\hat{\vec{p}}\times\hat{\vec{L}}-i\hbar\hat{\vec{p}})-k\frac{\vec{r}}{r}
\end{displaymath}
In the spherical coordinate,
\begin{displaymath}
\hat{A}_{x}\vec{i}+\hat{A}_{y}\vec{j}+\hat{A}_{z}\vec{k}=\hat{\vec{A}}=\frac{1}{\mu}(\hat{\vec{p}}\times\hat{\vec{L}}
-i\hbar\hat{\vec{p}})-k\frac{\vec{r}}{r}
=\left[
\begin{array}{ccc}
\vec{i} & \vec{j} & \vec{k}
\end{array}
\right]
\end{displaymath}
\begin{displaymath}
\left[
\begin{array}{c}
\frac{\hbar}{2\mu}(cos\theta \frac{\partial}{\partial r}-\frac{sin\theta}{r}\frac{\partial}{\partial\theta})(\hat{L}_{+}-\hat{L}_{-})-\frac{i\hbar}{\mu}
(sin\theta sin\varphi\frac{\partial}{\partial r}+\frac{cos\theta sin\varphi}{r}\frac{\partial}{\partial\theta}+\frac{cos\varphi}{rsin\theta}\frac{\partial}{\partial\varphi})\hat{L}_{z}\\
\frac{i\hbar}{\mu}(sin\theta cos\varphi\frac{\partial}{\partial r}+\frac{cos\theta cos\varphi}{r}\frac{\partial}{\partial\theta}-\frac{sin\varphi}{rsin\theta}\frac{\partial}{\partial\varphi})\hat{L}_{z}
-\frac{i\hbar}{2\mu}(cos\theta \frac{\partial}{\partial r}-\frac{sin\theta}{r}\frac{\partial}{\partial\theta})(\hat{L}_{+}+\hat{L}_{-})\\
\frac{\hbar e^{i\varphi}}{2\mu}(sin\theta\frac{\partial}{\partial r}+\frac{cos\theta}{r}\frac{\partial}{\partial \theta}+\frac{i}{rsin\theta}\frac{\partial}{\partial \varphi})\hat{L}_{-}-\frac{\hbar e^{-i\varphi}}{2\mu}(sin\theta\frac{\partial}{\partial r}+\frac{cos\theta}{r}\frac{\partial}{\partial\theta}-\frac{i}{rsin\theta}\frac{\partial}{\partial \varphi})\hat{L}_{+}\\
\end{array}
\right]
\end{displaymath}
\begin{displaymath}
-\left[
\begin{array}{ccc}
\vec{i} & \vec{j} & \vec{k}
\end{array}
\right]
\left[
\begin{array}{c}
\frac{\hbar^{2}}{\mu}(sin\theta cos\varphi \frac{\partial}{\partial r}+\frac{cos\theta cos\varphi}{r}\frac{\partial}{\partial\theta} -\frac{sin\varphi}{rsin\theta}\frac{\partial}{\partial \varphi})+ksin\theta cos\varphi\\
\frac{\hbar^{2}}{\mu}(sin\theta sin\varphi \frac{\partial}{\partial r}+\frac{cos\theta sin\varphi}{r}\frac{\partial}{\partial\theta} +\frac{cos\varphi}{rsin\theta}\frac{\partial}{\partial \varphi})+ksin\theta sin\varphi\\
\frac{\hbar^{2}}{\mu}(cos\theta\frac{\partial}{\partial r}-\frac{sin\theta}{r}\frac{\partial}{\partial \theta})+kcos\theta\\
\end{array}
\right]
\end{displaymath}
Let $\hat{A}_{+}=\hat{A}_{x}+i\hat{A}_{y}$ and $\hat{A}_{-}=\hat{A}_{x}-i\hat{A}_{y}$,
\begin{equation}
\left \{  \begin{array}{c}
\hat{A}_{+}=-\frac{\hbar e^{i\varphi}}{\mu}(sin\theta\frac{\partial}{\partial r} +\frac{cos\theta}{r}\frac{\partial}{\partial\theta}+\frac{i}{rsin\theta}\frac{\partial}{\partial\varphi})(\hat{L}_{z}+\hbar)\\
+\frac{\hbar}{\mu}(cos\theta\frac{\partial}{\partial r}-\frac{sin\theta}{r}\frac{\partial}{\partial\theta})\hat{L}_{+}-ksin\theta e^{i\varphi}  \\
\hat{A}_{-}=\frac{\hbar e^{-i\varphi}}{\mu}(sin\theta\frac{\partial}{\partial r}+\frac{cos\theta}{r}\frac{\partial}{\partial\theta}-\frac{i}{rsin\theta}\frac{\partial}{\partial\varphi})(\hat{L}_{z}-\hbar)\\
+\frac{\hbar}{\mu}(-cos\theta\frac{\partial}{\partial r}+\frac{sin\theta}{r}\frac{\partial}{\partial\theta})\hat{L}_{-}-ksin\theta e^{-i\varphi}  \\
\hat{A}_{z}=\frac{\hbar^{2}}{2\mu}[\frac{1}{\hbar}e^{i\varphi}(sin\theta\frac{\partial}{\partial r}+\frac{cos\theta}{r}\frac{\partial}{\partial\theta}
+\frac{i}{rsin\theta}\frac{\partial}{\partial\varphi})\hat{L}_{-}-\frac{1}{\hbar}e^{-i\varphi}\\
(sin\theta\frac{\partial}{\partial r}+\frac{cos\theta}{r}\frac{\partial}{\partial\theta}-\frac{i}{rsin\theta}
\frac{\partial}{\partial\varphi})\hat{L}_{+}-2(cos\theta\frac{\partial}{\partial r}-\frac{sin\theta}{r}\frac{\partial}{\partial\theta})]-kcos\theta
\end{array}\right.
\end{equation}
If the operators $A_{z}$, $A_{+}$ and $A_{-}$ in (42) act on the wave function $\psi_{nlm}$ in (41) respectively, then we will obtain
(88)-(90) which can be expanded to the matrix expressions. For example, when n=3,
\begin{displaymath}
\hat{A}_{z} \begin{array}{ccccccccc}[\psi_{300}&\psi_{31-1}&\psi_{310}&\psi_{311}&\psi_{322}&\psi_{321}&\psi_{320}&\psi_{32-1}&\psi_{32-2}]\end{array}
\end{displaymath}
\begin{displaymath}
=\begin{array}{ccccccccc}k[\frac{2\sqrt{6}}{9}\psi_{310}&\frac{\psi_{32-1}}{3}&\frac{2\sqrt{6}}{9}\psi_{300}+\frac{2\sqrt{3}}{9}\psi_{320}
&\frac{\psi_{321}}{3}&0&\frac{\psi_{311}}{3}&\frac{2\sqrt{3}}{9}\psi_{310}&\frac{\psi_{31-1}}{3}&0]\end{array}
\end{displaymath}
\begin{displaymath}
=\begin{array}{ccccccccc}[\psi_{300}&\psi_{31-1}&\psi_{310}&\psi_{311}&\psi_{322}&\psi_{321}&\psi_{320}&\psi_{32-1}&\psi_{32-2}]\end{array}
\end{displaymath}
\begin{displaymath}
\frac{k}{9}\left[
\begin{array}{ccccccccc}
0&0&2\sqrt{6}&0&0&0&0&0&0\\
0&0&0&0&0&0&0&3&0\\
2\sqrt{6}&0&0&0&0&0&2\sqrt{3}&0&0\\
0&0&0&0&0&3&0&0&0\\
0&0&0&0&0&0&0&0&0\\
0&0&0&3&0&0&0&0&0\\
0&0&2\sqrt{3}&0&0&0&0&0&0\\
0&3&0&0&0&0&0&0&0\\
0&0&0&0&0&0&0&0&0\\
\end{array}
\right]
\end{displaymath}
So we can see that operator $\hat{A}_{z}$($\hat{A}_{+}$ or $\hat{A}_{-}$), square matrix and wave functions which are the solutions of second order Schr$\ddot{o}$dinger
equation and the orthonormalized simultaneous eigenfunctions of the operators $\hat{H},\hat{L}_{z}$ and $\hat{L}^{2}$ are represented in the same expression.\\\\
Scheme II\\
We can assume that there are some following sets of orthonormalized wave functions\\
When $n$=1\\
\begin{displaymath}
\psi_{100}=R_{10}Y_{00}
\end{displaymath}
When $n$=2\\
\begin{displaymath}
\psi_{200}=R_{20}Y_{00};\psi_{21-1}=R_{21}Y_{1-1},\psi_{210}=R_{21}Y_{10},\psi_{211}=R_{21}Y_{11}
\end{displaymath}
$\cdots,\cdots,$\\
When $n$=$s\rightarrow \infty$\\
\begin{displaymath}
\psi_{s00}=R_{s0}Y_{00};\psi_{s1-1}=R_{s1}Y_{1-1},\psi_{s10}=R_{s1}Y_{10},\psi_{s11}=R_{s1}Y_{11};\cdots;
\end{displaymath}
\begin{small}
	\begin{displaymath}
	\psi_{ss-1s-1}=R_{ss-1}Y_{s-1s-1},\cdots,\psi_{ss-12-s}=R_{ss-1}Y_{s-12-s},\psi_{ss-11-s}=R_{ss-1}Y_{s-11-s}
	\end{displaymath}
\end{small}
where the radial wave functions are unknown and s is an odd number. When n is odd, we must arrange the wave functions in the row matrix as follows
\begin{displaymath}
\hat{L}_{z} \begin{array}{ccccccccc}[\psi_{300}&\psi_{31-1}&\psi_{310}&\psi_{311}&\psi_{322}&\psi_{321}&\psi_{320}&\psi_{32-1}&\psi_{32-2}]\end{array}
\end{displaymath}
\begin{displaymath}
=\begin{array}{ccccccccc}[\psi_{300}&\psi_{31-1}&\psi_{310}&\psi_{311}&\psi_{322}&\psi_{321}&\psi_{320}&\psi_{32-1}&\psi_{32-2}]\end{array}
\end{displaymath}
\begin{displaymath}
\hbar \left[
\begin{array}{ccccccccc}
0&0&0&0&0&0&0&0&0\\
0&-1&0&0&0&0&0&0&0\\
0&0&0&0&0&0&0&0&0\\
0&0&0&1&0&0&0&0&0\\
0&0&0&0&2&0&0&0&0\\
0&0&0&0&0&1&0&0&0\\
0&0&1&0&0&0&0&0&0\\
0&0&0&0&0&0&0&-1&0\\
0&0&0&0&0&0&0&0&-2\\
\end{array}
\right]
\end{displaymath}
We have to ensure that the eigenvalue of $\hat{L}_{z}$ raises or lowers $\hbar$, rather than jumping dropping by more than 2$\hbar$ so that it can
make the system self-consistent. The following expression is not allowed.
\begin{displaymath}
\hat{L}_{z} \begin{array}{ccccccccc}[\psi_{300}&\psi_{31-1}&\psi_{310}&\psi_{311}&\psi_{32-2}&\psi_{32-1}&\psi_{320}&\psi_{321}&\psi_{322}]\end{array}
\end{displaymath}
\begin{displaymath}
=\begin{array}{ccccccccc}[\psi_{300}&\psi_{31-1}&\psi_{310}&\psi_{311}&\psi_{32-2}&\psi_{32-1}&\psi_{320}&\psi_{321}&\psi_{322}]\end{array}
\end{displaymath}
\begin{displaymath}
\hbar \left[
\begin{array}{ccccccccc}
0&0&0&0&0&0&0&0&0\\
0&-1&0&0&0&0&0&0&0\\
0&0&0&0&0&0&0&0&0\\
0&0&0&1&0&0&0&0&0\\
0&0&0&0&-2&0&0&0&0\\
0&0&0&0&0&-1&0&0&0\\
0&0&1&0&0&0&0&0&0\\
0&0&0&0&0&0&0&1&0\\
0&0&0&0&0&0&0&0&2\\
\end{array}
\right]
\end{displaymath}
We can show that
\begin{equation}
[\hat{L}_{z},\hat{A}_{z}]=0
\end{equation}
\begin{equation}
\hbar\hat{A}_{+}=[\hat{A}_{z},\hat{L}_{+}]
\end{equation}
\begin{equation}
2\hbar\hat{A}_{z}=[\hat{A}_{+},\hat{L}_{-}]
\end{equation}
\begin{equation}
\hbar\hat{A}_{-}=[\hat{L}_{-},\hat{A}_{z}]
\end{equation}
\begin{equation}
\hat{A}_{+}\hat{L}_{-}+\hat{A}_{-}\hat{L}_{+}+2\hat{A}_{z}\hat{L}_{z}=0
\end{equation}
\begin{equation}
2\hat{H}(\hat{L}^2+\hbar\hat{L}_{z}+\hbar^2)+\mu k^{2}=\mu (\hat{A}_{-}\hat{A}_{+}+\hat{A}_{z}^2)
\end{equation}
\begin{equation}
\mu[\hat{A}_{-},\hat{A}_{+}]=4\hbar\hat{H}\hat{L}_{z}
\end{equation}
\textbf{When n=1}\\
From (29)-(32),
\begin{equation}
\hat{L}_{z}\psi_{100}=\hat{L}_{+}\psi_{100}=\hat{L}_{-}\psi_{100}=\hat{L}^{2}\psi_{100}=0
\end{equation}
Because $\hat{A}_{z}$ is a Hermitian operator, it is assumed that
\begin{equation}
\hat{A}_{z}\psi_{100}=A\psi_{100}
\end{equation}
where $A$ are real numbers. Combining with (44) and (50),
\begin{equation}
\hat{A}_{+}\psi_{100}=\frac{[\hat{A}_{z},\hat{L}_{+}]\psi_{100}}{\hbar}=\frac{(\hat{A}_{z}\hat{L}_{+}\psi_{100}-\hat{L}_{+}\hat{A}_{z}\psi_{100})}{\hbar}
=-\frac{A}{\hbar}\hat{L}_{+}\psi_{100}=0
\end{equation}
From (45), (50) and (52),
\begin{equation}
\hat{A}_{z}\psi_{100}=\frac{[\hat{A}_{+},\hat{L}_{-}]\psi_{100}}{2\hbar}=\frac{(\hat{A}_{+}\hat{L}_{-}\psi_{100}-\hat{L}_{-}\hat{A}_{+}\psi_{100})}{2\hbar}=0
\end{equation}
From (46), (50) and (53),
\begin{equation}
\hat{A}_{-}\psi_{100}=\frac{[\hat{L}_{-},\hat{A}_{z}]\psi_{100}}{\hbar}=\frac{(\hat{L}_{-}\hat{A}_{z}\psi_{100}-\hat{A}_{z}\hat{L}_{-}\psi_{100})}{\hbar}=0
\end{equation}
And that from (38), (48), (50) and (52)-(54),
\begin{equation}
(2\hbar^{2}E_{1}+\mu k^{2})\psi_{100}=0 \Rightarrow E_{1}=-\frac{\mu k^2}{2\hbar^2}
\end{equation}
\textbf{When n=2:}\\
From (29)-(32),
\begin{equation}
\hat{L}_{z} \setlength\arraycolsep{0.2em} \begin{array}{cccc}[\psi_{200}&\psi_{21-1}&\psi_{210}&\psi_{211}]\end{array}
=\setlength\arraycolsep{0.2em} \begin{array}{cccc}[\psi_{200}&\psi_{21-1}&\psi_{210}&\psi_{211}]\end{array}
\hbar\left[\setlength\arraycolsep{0.3em}
\begin{array}{cccc}
0&0&0&0\\
0&-1&0&0\\
0&0&0&0\\
0&0&0&1
\end{array}
\right]
\end{equation}
\begin{equation}
\hat{L}_{-} \setlength\arraycolsep{0.2em} \begin{array}{cccc}[\psi_{200}&\psi_{21-1}&\psi_{210}&\psi_{211}]\end{array}
=\setlength\arraycolsep{0.2em} \begin{array}{cccc}[\psi_{200}&\psi_{21-1}&\psi_{210}&\psi_{211}]\end{array}
\sqrt{2}\hbar\left[\setlength\arraycolsep{0.3em}
\begin{array}{cccc}
0&0&0&0\\
0&0&1&0\\
0&0&0&1\\
0&0&0&0
\end{array}
\right]
\end{equation}
\begin{equation}
\hat{L}_{+} \setlength\arraycolsep{0.2em} \begin{array}{cccc}[\psi_{200}&\psi_{21-1}&\psi_{210}&\psi_{211}]\end{array}
=\setlength\arraycolsep{0.2em} \begin{array}{cccc}[\psi_{200}&\psi_{21-1}&\psi_{210}&\psi_{211}]\end{array}
\sqrt{2}\hbar\left[\setlength\arraycolsep{0.3em}
\begin{array}{cccc}
0&0&0&0\\
0&0&0&0\\
0&1&0&0\\
0&0&1&0
\end{array}
\right]
\end{equation}
\begin{equation}
\setlength\arraycolsep{0.2em} \hat{L}^{2}\begin{array}{cccc}[\psi_{200}&\psi_{21-1}&\psi_{210}&\psi_{211}]\end{array}
=\setlength\arraycolsep{0.2em} \begin{array}{cccc}[\psi_{200}&\psi_{21-1}&\psi_{210}&\psi_{211}]\end{array}
\hbar^{2}\left[ \setlength\arraycolsep{0.3em}
\begin{array}{cccc}
0&0&0&0\\
0&2&0&0\\
0&0&2&0\\
0&0&0&2
\end{array}
\right]
\end{equation}
Because $\hat{A}_{z}$ is a Hermitian operator, it is assumed that:
\begin{displaymath}
\hat{A}_{z}\begin{array}{cccc}[\psi_{200}&\psi_{21-1}&\psi_{210}&\psi_{211}]\end{array}
=\begin{array}{cccc}[\psi_{200}&\psi_{21-1}&\psi_{210}&\psi_{211}]\end{array}
\end{displaymath}
\begin{equation}
\left[
\begin{array}{cccc}
A_{11}(2)&A^{*}_{21}(2)&A^{*}_{31}(2)&A^{*}_{41}(2)\\
A_{21}(2)&A_{22}(2)&A^{*}_{32}(2)&A^{*}_{42}(2)\\
A_{31}(2)&A_{32}(2)&A_{33}(2)&A^{*}_{43}(2)\\
A_{41}(2)&A_{42}(2)&A_{43}(2)&A_{44}(2)
\end{array}
\right]
\end{equation}
where $A_{11}(2)$, $A_{22}(2)$, $A_{33}(2)$ and $A_{44}(2)$ are real numbers. From (43), (56) and (60),
\begin{displaymath}
\hat{L}_{z}\hat{A}_{z}\begin{array}{cccc}[\psi_{200}&\psi_{21-1}&\psi_{210}&\psi_{211}]\end{array}
=\hat{A}_{z}\hat{L}_{z}\begin{array}{cccc}[\psi_{200}&\psi_{21-1}&\psi_{210}&\psi_{211}]\end{array}
\end{displaymath}
So
\begin{displaymath}
A_{21}=A_{41}=A_{32}=A_{42}=A_{43}=0
\end{displaymath}
Thus
\begin{displaymath}
\hat{A}_{z}\begin{array}{cccc}[\psi_{200}&\psi_{21-1}&\psi_{210}&\psi_{211}]\end{array}
=\begin{array}{cccc}[\psi_{200}&\psi_{21-1}&\psi_{210}&\psi_{211}]\end{array}
\end{displaymath}
\begin{equation}
\left[
\begin{array}{cccc}
A_{11}(2)&0&A^{*}_{31}(2)&0\\
0&A_{22}(2)&0&0\\
A_{31}(2)&0&A_{33}(2)&0\\
0&0&0&A_{44}(2)
\end{array}
\right]
\end{equation}
Combining with (44) and (58),
\begin{displaymath}
\hat{A}_{+}\begin{array}{cccc}[\psi_{200}&\psi_{21-1}&\psi_{210}&\psi_{211}]\end{array}
=\begin{array}{cccc}[\psi_{200}&\psi_{21-1}&\psi_{210}&\psi_{211}]\end{array}
\end{displaymath}
\begin{equation}
\sqrt{2}\left[
\begin{array}{cccc}
0&A^{*}_{31}(2)&0&0\\
0&0&0&0\\
0&A_{33}(2)-A_{22}(2)&0&0\\
-A_{31}(2)&0&A_{44}(2)-A_{33}(2)&0
\end{array}
\right]
\end{equation}
From (45),
\begin{displaymath}
2\hbar \hat{A}_{z}\begin{array}{cccc}[\psi_{200}&\psi_{21-1}&\psi_{210}&\psi_{211}]\end{array}
=[\hat{A}_{+},\hat{L}_{-}]\begin{array}{cccc}[\psi_{200}&\psi_{21-1}&\psi_{210}&\psi_{211}]\end{array}
\end{displaymath}
Combining with (57) and (61)-(62),
\begin{equation}
A_{33}(2)=A_{11}(2)=0,A_{22}(2)=-A_{44}(2)
\end{equation}
Combining with (46), (57) and (61),
\begin{displaymath}
\hat{A}_{-}\begin{array}{cccc}[\psi_{200}&\psi_{21-1}&\psi_{210}&\psi_{211}]\end{array}
=\begin{array}{cccc}[\psi_{200}&\psi_{21-1}&\psi_{210}&\psi_{211}]\end{array}
\end{displaymath}
\begin{equation}
\sqrt{2}\left[
\begin{array}{cccc}
0&0&0&-A^{*}_{31}(2)\\
A_{31}(2)&0&A_{44}(2)&0\\
0&0&0&A_{44}(2)\\
0&0&0&0
\end{array}
\right]
\end{equation}
From (47), (56)-(58) and (61)-(64),
\begin{small}
	\begin{displaymath}
	(\hat{A}_{+}\hat{L}_{-}+\hat{A}_{-}\hat{L}_{+}+2\hat{A}_{z}\hat{L}_{z})\psi_{211}=\hbar(\sqrt{2}\hat{A}_{+}\psi_{210}+2\hat{A}_{z}\psi_{211})
	=4\hbar A_{44}(2)\psi_{211}=0
	\end{displaymath}
\end{small}
So
\begin{displaymath}
A_{44}(2)=0
\end{displaymath}
Therefore, the equations (61)-(62) and (64) become respectively
\begin{displaymath}
\hat{A}_{z}\begin{array}{cccc}[\psi_{200}&\psi_{21-1}&\psi_{210}&\psi_{211}]\end{array}
=\begin{array}{cccc}[\psi_{200}&\psi_{21-1}&\psi_{210}&\psi_{211}]\end{array}
\end{displaymath}
\begin{equation}
\left[
\begin{array}{cccc}
0&0&A^{*}_{31}(2)&0\\
0&0&0&0\\
A_{31}(2)&0&0&0\\
0&0&0&0
\end{array}
\right]
\end{equation}
\begin{displaymath}
\hat{A}_{+}\begin{array}{cccc}[\psi_{200}&\psi_{21-1}&\psi_{210}&\psi_{211}]\end{array}
=\begin{array}{cccc}[\psi_{200}&\psi_{21-1}&\psi_{210}&\psi_{211}]\end{array}
\end{displaymath}
\begin{equation}
\sqrt{2}\left[
\begin{array}{cccc}
0&A^{*}_{31}(2)&0&0\\
0&0&0&0\\
0&0&0&0\\
-A_{31}(2)&0&0&0
\end{array}
\right]
\end{equation}
\begin{displaymath}
\hat{A}_{-}\begin{array}{cccc}[\psi_{200}&\psi_{21-1}&\psi_{210}&\psi_{211}]\end{array}
=\begin{array}{cccc}[\psi_{200}&\psi_{21-1}&\psi_{210}&\psi_{211}]\end{array}
\end{displaymath}
\begin{equation}
\sqrt{2}\left[
\begin{array}{cccc}
0&0&0&-A^{*}_{31}(2)\\
A_{31}(2)&0&0&0\\
0&0&0&0\\
0&0&0&0
\end{array}
\right]
\end{equation}
Combining with (48), (56) and (59),
\begin{displaymath}
(8\hbar^{2}E_{2}+\mu k^{2})\psi_{211}=[2\hat{H}(\hat{L}^{2}+\hbar\hat{L}_{z}+\hbar^{2})+\mu k^{2}]\psi_{211}
=\mu(\hat{A}_{-}\hat{A}_{+}+\hat{A}_{z}^{2})\psi_{211}=0
\end{displaymath}
So
\begin{equation}
E_{2}=-\frac{\mu k^2}{2\hbar^2}\frac{1}{2^2}
\end{equation}
From (49), (56) and (66)-(67),
\begin{small}
	\begin{displaymath}
	-2\mu |A_{31}(2)|^{2}\psi_{211}\!=\!\sqrt{2}\mu A^{*}_{31}(2)\hat{A}_{+}\psi_{200}\!=\!\mu [\hat{A}_{-}, \hat{A}_{+}]\psi_{211}
	\!=\!4\hbar \hat{H}\hat{L}_{z}\psi_{211}\!=\!4\hbar^{2}E_{2}\psi_{211}
	\end{displaymath}
\end{small}
Thus
\begin{equation}
|A_{31}(2)|^2=\frac{k^2}{4}
\end{equation}
If we take positive real solutions from (69), then (65)-(67) become
\begin{equation}
\left \{  \begin{array}{c}
\hat{A}_{z}\psi_{200}=\frac{k}{2}\psi_{210} \\
\hat{A}_{z}\psi_{21-1}=0,\hat{A}_{z}\psi_{210}=\frac{k}{2}\psi_{200},\hat{A}_{z}\psi_{211}=0
\end{array}\right.
\end{equation}
\begin{equation}
\left \{  \begin{array}{c}
\hat{A}_{+}\psi_{200}=-\frac{\sqrt{2}}{2}k\psi_{211} \\
\hat{A}_{+}\psi_{21-1}=\frac{\sqrt{2}}{2}k\psi_{200},\hat{A}_{+}\psi_{210}=0,\hat{A}_{+}\psi_{211}=0
\end{array}\right.
\end{equation}
\begin{equation}
\left \{  \begin{array}{c}
\hat{A}_{-}\psi_{200}=\frac{\sqrt{2}}{2}k\psi_{21-1} \\
\hat{A}_{-}\psi_{21-1}=0,\hat{A}_{-}\psi_{210}=0,\hat{A}_{-}\psi_{211}=-\frac{\sqrt{2}}{2}k\psi_{200}
\end{array}\right.
\end{equation}
$\cdots,\cdots,$\\\\
\textbf{When $n=s\rightarrow\infty$}\\\\
From (29)-(32),
\begin{tiny}
	\begin{displaymath}\setlength\arraycolsep{0.1em}
	\begin{array}{cccccccccccccccc}\hat{L}_{z}[\psi_{\!s0\!0\!}&\psi_{\!s1\!-\!1\!}&\psi_{\!s10}&\cdots&\psi_{\!ss\!-\!33\!-\!s}&\psi_{\!ss\!-\!22\!-\!s}
	&\psi_{\!ss\!-\!23\!-\!s}&\cdots&\psi_{\!ss\!-\!2s\!-\!2}&\psi_{\!ss\!-\!1s\!-\!1}&\psi_{\!ss\!-\!1s\!-\!2}&\psi_{\!ss\!-\!1s\!-\!3}&\cdots
	&\psi_{\!ss\!-\!13\!-\!s}&\psi_{\!ss\!-\!12\!-\!s}&\psi_{\!ss\!-\!11\!-\!s}] \end{array}
	\end{displaymath}
\end{tiny}
\begin{tiny}
	\begin{displaymath}\setlength\arraycolsep{0.1em}
	=\begin{array}{cccccccccccccccc}[\psi_{\!s0\!0\!}&\psi_{\!s1\!-\!1\!}&\psi_{\!s10}&\cdots&\psi_{\!ss\!-\!33\!-\!s}&\psi_{\!ss\!-\!22\!-\!s}
	&\psi_{\!ss\!-\!23\!-\!s}&\cdots&\psi_{\!ss\!-\!2s\!-\!2}&\psi_{\!ss\!-\!1s\!-\!1}&\psi_{\!ss\!-\!1s\!-\!2}&\psi_{\!ss\!-\!1s\!-\!3}&\cdots
	&\psi_{\!ss\!-\!13\!-\!s}&\psi_{\!ss\!-\!12\!-\!s}&\psi_{\!ss\!-\!11\!-\!s}]\hbar \end{array}
	\end{displaymath}
\end{tiny}
\begin{small}
	\begin{equation}
	\left[\setlength\arraycolsep{0em}
	\begin{array}{cccccccccccccccc}
	0&0&0&\cdots&0&0&0&\cdots&0&0&0&0&\cdots&0&0&0\\
	0&-1&0&\cdots&0&0&0&\cdots&0&0&0&0&\cdots&0&0&0\\
	0&0&0&\cdots&0&0&0&\cdots&0&0&0&0&\cdots&0&0&0\\
	\cdots&\cdots&\cdots&\cdots&\cdots&\cdots&\cdots&\cdots&\cdots&\cdots&\cdots&\cdots&\cdots&\cdots&\cdots&\cdots\\
	0&0&0&\cdots&3-s&0&0&\cdots&0&0&0&0&\cdots&0&0&0\\
	0&0&0&\cdots&0&2-s&0&\cdots&0&0&0&0&\cdots&0&0&0\\
	0&0&0&\cdots&0&0&3-s&\cdots&0&0&0&0&\cdots&0&0&0\\
	\cdots&\cdots&\cdots&\cdots&\cdots&\cdots&\cdots&\cdots&\cdots&\cdots&\cdots&\cdots&\cdots&\cdots&\cdots&\cdots\\
	0&0&0&\cdots&0&0&0&\cdots&s-2&0&0&0&\cdots&0&0&0\\
	0&0&0&\cdots&0&0&0&\cdots&0&s-1&0&0&\cdots&0&0&0\\
	0&0&0&\cdots&0&0&0&\cdots&0&0&s-2&0&\cdots&0&0&0\\
	0&0&0&\cdots&0&0&0&\cdots&0&0&0&s-3&\cdots&0&0&0\\
	\cdots&\cdots&\cdots&\cdots&\cdots&\cdots&\cdots&\cdots&\cdots&\cdots&\cdots&\cdots&\cdots&\cdots&\cdots&\cdots\\
	0&0&0&\cdots&0&0&0&\cdots&0&0&0&0&\cdots&3-s&0&0\\
	0&0&0&\cdots&0&0&0&\cdots&0&0&0&0&\cdots&0&2-s&0\\
	0&0&0&\cdots&0&0&0&\cdots&0&0&0&0&\cdots&0&0&1-s\\
	\end{array}
	\right]
	\end{equation}
\end{small}
\begin{tiny}
	\begin{displaymath}\setlength\arraycolsep{0.1em}
	\begin{array}{cccccccccccccccc}\hat{L}_{-}[\psi_{\!s0\!0\!}&\psi_{\!s1\!-\!1\!}&\psi_{\!s10}&\cdots&\psi_{\!ss\!-\!33\!-\!s}&\psi_{\!ss\!-\!22\!-\!s}
	&\psi_{\!ss\!-\!23\!-\!s}&\cdots&\psi_{\!ss\!-\!2s\!-\!2}&\psi_{\!ss\!-\!1s\!-\!1}&\psi_{\!ss\!-\!1s\!-\!2}&\psi_{\!ss\!-\!1s\!-\!3}&\cdots
	&\psi_{\!ss\!-\!13\!-\!s}&\psi_{\!ss\!-\!12\!-\!s}&\psi_{\!ss\!-\!11\!-\!s}] \end{array}
	\end{displaymath}
\end{tiny}
\begin{tiny}
	\begin{displaymath}\setlength\arraycolsep{0.1em}
	=\begin{array}{cccccccccccccccc}[\psi_{\!s0\!0\!}&\psi_{\!s1\!-\!1\!}&\psi_{\!s10}&\cdots&\psi_{\!ss\!-\!33\!-\!s}&\psi_{\!ss\!-\!22\!-\!s}
	&\psi_{\!ss\!-\!23\!-\!s}&\cdots&\psi_{\!ss\!-\!2s\!-\!2}&\psi_{\!ss\!-\!1s\!-\!1}&\psi_{\!ss\!-\!1s\!-\!2}&\psi_{\!ss\!-\!1s\!-\!3}&\cdots
	&\psi_{\!ss\!-\!13\!-\!s}&\psi_{\!ss\!-\!12\!-\!s}&\psi_{\!ss\!-\!11\!-\!s}]\hbar \end{array}
	\end{displaymath}
\end{tiny}
\begin{scriptsize}
	\begin{equation}
	\left[\setlength\arraycolsep{0em}
	\begin{array}{cccccccccccccccc}
	0&0&0&\cdots&0&0&0&\cdots&0&0&0&0&\cdots&0&0&0\\
	0&0&\sqrt{2}&\cdots&0&0&0&\cdots&0&0&0&0&\cdots&0&0&0\\
	0&0&0&\cdots&0&0&0&\cdots&0&0&0&0&\cdots&0&0&0\\
	\cdots&\cdots&\cdots&\cdots&\cdots&\cdots&\cdots&\cdots&\cdots&\cdots&\cdots&\cdots&\cdots&\cdots&\cdots&\cdots\\
	0&0&0&\cdots&0&0&0&\cdots&0&0&0&0&\cdots&0&0&0\\
	0&0&0&\cdots&0&0&\sqrt{2s-4}&\cdots&0&0&0&0&\cdots&0&0&0\\
	0&0&0&\cdots&0&0&0&\cdots&0&0&0&0&\cdots&0&0&0\\
	\cdots&\cdots&\cdots&\cdots&\cdots&\cdots&\cdots&\cdots&\cdots&\cdots&\cdots&\cdots&\cdots&\cdots&\cdots&\cdots\\
	0&0&0&\cdots&0&0&0&\cdots&0&0&0&0&\cdots&0&0&0\\
	0&0&0&\cdots&0&0&0&\cdots&0&0&0&0&\cdots&0&0&0\\
	0&0&0&\cdots&0&0&0&\cdots&0&\sqrt{2s-2}&0&0&\cdots&0&0&0\\
	0&0&0&\cdots&0&0&0&\cdots&0&0&\sqrt{2(2s-3)}&0&\cdots&0&0&0\\
	\cdots&\cdots&\cdots&\cdots&\cdots&\cdots&\cdots&\cdots&\cdots&\cdots&\cdots&\cdots&\cdots&\cdots&\cdots&\cdots\\
	0&0&0&\cdots&0&0&0&\cdots&0&0&0&0&\cdots&0&0&0\\
	0&0&0&\cdots&0&0&0&\cdots&0&0&0&0&\cdots&\sqrt{(2s-3)2}&0&0\\
	0&0&0&\cdots&0&0&0&\cdots&0&0&0&0&\cdots&0&\sqrt{2s-2}&0\\
	\end{array}
	\right]
	\end{equation}
\end{scriptsize}
\begin{tiny}
	\begin{displaymath}\setlength\arraycolsep{0.1em}
	\begin{array}{cccccccccccccccc}\hat{L}_{+}[\psi_{\!s0\!0\!}&\psi_{\!s1\!-\!1\!}&\psi_{\!s10}&\cdots&\psi_{\!ss\!-\!33\!-\!s}&\psi_{\!ss\!-\!22\!-\!s}
	&\psi_{\!ss\!-\!23\!-\!s}&\cdots&\psi_{\!ss\!-\!2s\!-\!2}&\psi_{\!ss\!-\!1s\!-\!1}&\psi_{\!ss\!-\!1s\!-\!2}&\psi_{\!ss\!-\!1s\!-\!3}&\cdots
	&\psi_{\!ss\!-\!13\!-\!s}&\psi_{\!ss\!-\!12\!-\!s}&\psi_{\!ss\!-\!11\!-\!s}] \end{array}
	\end{displaymath}
\end{tiny}
\begin{tiny}
	\begin{displaymath}\setlength\arraycolsep{0.1em}
	=\begin{array}{cccccccccccccccc}[\psi_{\!s0\!0\!}&\psi_{\!s1\!-\!1\!}&\psi_{\!s10}&\cdots&\psi_{\!ss\!-\!33\!-\!s}&\psi_{\!ss\!-\!22\!-\!s}
	&\psi_{\!ss\!-\!23\!-\!s}&\cdots&\psi_{\!ss\!-\!2s\!-\!2}&\psi_{\!ss\!-\!1s\!-\!1}&\psi_{\!ss\!-\!1s\!-\!2}&\psi_{\!ss\!-\!1s\!-\!3}&\cdots
	&\psi_{\!ss\!-\!13\!-\!s}&\psi_{\!ss\!-\!12\!-\!s}&\psi_{\!ss\!-\!11\!-\!s}]\hbar \end{array}
	\end{displaymath}
\end{tiny}
\begin{scriptsize}
	\begin{equation}
	\left[\setlength\arraycolsep{0em}
	\begin{array}{cccccccccccccccc}
	0&0&0&\cdots&0&0&0&\cdots&0&0&0&0&\cdots&0&0&0\\
	0&0&0&\cdots&0&0&0&\cdots&0&0&0&0&\cdots&0&0&0\\
	0&\sqrt{2}&0&\cdots&0&0&0&\cdots&0&0&0&0&\cdots&0&0&0\\
	\cdots&\cdots&\cdots&\cdots&\cdots&\cdots&\cdots&\cdots&\cdots&\cdots&\cdots&\cdots&\cdots&\cdots&\cdots&\cdots\\
	0&0&0&\cdots&0&0&0&\cdots&0&0&0&0&\cdots&0&0&0\\
	0&0&0&\cdots&0&0&0&\cdots&0&0&0&0&\cdots&0&0&0\\
	0&0&0&\cdots&0&\sqrt{2s-4}&0&\cdots&0&0&0&0&\cdots&0&0&0\\
	\cdots&\cdots&\cdots&\cdots&\cdots&\cdots&\cdots&\cdots&\cdots&\cdots&\cdots&\cdots&\cdots&\cdots&\cdots&\cdots\\
	0&0&0&\cdots&0&0&0&\cdots&0&0&0&0&\cdots&0&0&0\\
	0&0&0&\cdots&0&0&0&\cdots&0&0&\sqrt{2s-2}&0&\cdots&0&0&0\\
	0&0&0&\cdots&0&0&0&\cdots&0&0&0&\sqrt{2(2s-3)}&\cdots&0&0&0\\
	0&0&0&\cdots&0&0&0&\cdots&0&0&0&0&\cdots&0&0&0\\
	\cdots&\cdots&\cdots&\cdots&\cdots&\cdots&\cdots&\cdots&\cdots&\cdots&\cdots&\cdots&\cdots&\cdots&\cdots&\cdots\\
	0&0&0&\cdots&0&0&0&\cdots&0&0&0&0&\cdots&0&\sqrt{(2s-3)2}&0\\
	0&0&0&\cdots&0&0&0&\cdots&0&0&0&0&\cdots&0&0&\sqrt{2s-2}\\
	0&0&0&\cdots&0&0&0&\cdots&0&0&0&0&\cdots&0&0&0\\
	\end{array}
	\right]
	\end{equation}
\end{scriptsize}
\begin{tiny}
	\begin{displaymath}\setlength\arraycolsep{0.1em}
	\begin{array}{cccccccccccccccc}\hat{L}^{2}[\psi_{\!s0\!0\!}&\psi_{\!s1\!-\!1\!}&\psi_{\!s10}&\cdots&\psi_{\!ss\!-\!33\!-\!s}&\psi_{\!ss\!-\!22\!-\!s}
	&\psi_{\!ss\!-\!23\!-\!s}&\cdots&\psi_{\!ss\!-\!2s\!-\!2}&\psi_{\!ss\!-\!1s\!-\!1}&\psi_{\!ss\!-\!1s\!-\!2}&\psi_{\!ss\!-\!1s\!-\!3}&\cdots
	&\psi_{\!ss\!-\!13\!-\!s}&\psi_{\!ss\!-\!12\!-\!s}&\psi_{\!ss\!-\!11\!-\!s}] \end{array}
	\end{displaymath}
\end{tiny}
\begin{tiny}
	\begin{displaymath}\setlength\arraycolsep{0.1em}
	=\begin{array}{cccccccccccccccc}[\psi_{\!s0\!0\!}&\psi_{\!s1\!-\!1\!}&\psi_{\!s10}&\cdots&\psi_{\!ss\!-\!33\!-\!s}&\psi_{\!ss\!-\!22\!-\!s}
	&\psi_{\!ss\!-\!23\!-\!s}&\cdots&\psi_{\!ss\!-\!2s\!-\!2}&\psi_{\!ss\!-\!1s\!-\!1}&\psi_{\!ss\!-\!1s\!-\!2}&\psi_{\!ss\!-\!1s\!-\!3}&\cdots
	&\psi_{\!ss\!-\!13\!-\!s}&\psi_{\!ss\!-\!12\!-\!s}&\psi_{\!ss\!-\!11\!-\!s}]\hbar^{2} \end{array}
	\end{displaymath}
\end{tiny}
\begin{scriptsize}
	\begin{equation}
	\left[\setlength\arraycolsep{0em}
	\begin{array}{cccccccccccccccc}
	0&0&0&\cdots&0&0&0&\cdots&0&0&0&0&\cdots&0&0&0\\
	0&2&0&\cdots&0&0&0&\cdots&0&0&0&0&\cdots&0&0&0\\
	0&0&2&\cdots&0&0&0&\cdots&0&0&0&0&\cdots&0&0&0\\
	\cdots&\cdots&\cdots&\cdots&\cdots&\cdots&\cdots&\cdots&\cdots&\cdots&\cdots&\cdots&\cdots&\cdots&\cdots&\cdots\\
	0&0&0&\cdots&\!(\!s\!-\!3\!)(\!s\!-\!2\!)\!&0&0&\cdots&0&0&0&0&\cdots&0&0&0\\
	0&0&0&\cdots&0&\!(\!s\!-\!2\!)(\!s\!-\!1\!)\!&0&\cdots&0&0&0&0&\cdots&0&0&0\\
	0&0&0&\cdots&0&0&\!(\!s\!-\!2\!)(\!s\!-\!1\!)\!&\cdots&0&0&0&0&\cdots&0&0&0\\
	\cdots&\cdots&\cdots&\cdots&\cdots&\cdots&\cdots&\cdots&\cdots&\cdots&\cdots&\cdots&\cdots&\cdots&\cdots&\cdots\\
	0&0&0&\cdots&0&0&0&\cdots&\!(\!s\!-\!2\!)(\!s\!-\!1\!)\!&0&0&0&\cdots&0&0&0\\
	0&0&0&\cdots&0&0&0&\cdots&0&\!(\!s\!-\!1\!)s\!&0&0&\cdots&0&0&0\\
	0&0&0&\cdots&0&0&0&\cdots&0&0&\!(\!s\!-\!1\!)s\!&0&\cdots&0&0&0\\
	0&0&0&\cdots&0&0&0&\cdots&0&0&0&\!(\!s\!-\!1\!)s\!&\cdots&0&0&0\\
	\cdots&\cdots&\cdots&\cdots&\cdots&\cdots&\cdots&\cdots&\cdots&\cdots&\cdots&\cdots&\cdots&\cdots&\cdots&\cdots\\
	0&0&0&\cdots&0&0&0&\cdots&0&0&0&0&\cdots&\!(\!s\!-\!1\!)s\!&0&0\\
	0&0&0&\cdots&0&0&0&\cdots&0&0&0&0&\cdots&0&\!(\!s\!-\!1\!)s\!&0\\
	0&0&0&\cdots&0&0&0&\cdots&0&0&0&0&\cdots&0&0&\!(\!s\!-\!1\!)s\!\\
	\end{array}
	\right]
	\end{equation}
\end{scriptsize}\\
Because $\hat{A}_{z}$ is a Hermitian operator, it is assumed in terms of the theorem that:
\begin{displaymath}
\hat{A_{z}}\begin{array}{ccccccccc}[\psi_{s00}&\cdots&\psi_{ss-33-s}&\psi_{ss-22-s}&\cdots&\psi_{ss-2s-2}&\psi_{ss-1s-1}&\cdots&\psi_{ss-11-s}]\end{array}
\end{displaymath}
\begin{displaymath}
=\begin{array}{ccccccccc}[\psi_{s00}&\cdots&\psi_{ss-33-s}&\psi_{ss-22-s}&\cdots&\psi_{ss-2s-2}&\psi_{ss-1s-1}&\cdots&\psi_{ss-11-s}]\end{array}
\end{displaymath}
\begin{tiny}
	\begin{equation}
	\left[\setlength\arraycolsep{0em}
	\begin{array}{ccccccccc}
	A_{11}(s)&\cdots&A^{*}_{(s-2)^{2}1}(s)&A^{*}_{(s-2)^{2}+11}(s)&\cdots&A^{*}_{(s-1)^{2}1}(s)&A^{*}_{(s-1)^{2}+11}(s)&\cdots&A^{*}_{s^{2}1}(s)\\
	\cdots&\cdots&\cdots&\cdots&\cdots&\cdots&\cdots&\cdots&\cdots\\
	A_{\!(\!s\!-\!2\!)^{2}1}(\!s\!)&\cdots&A_{(s-2)^{2}(s-2)^{2}}(\!s\!)&A^{*}_{(s-2)^{2}+1(s-2)^{2}}(\!s\!)&\cdots&A^{*}_{(s-1)^{2}(s-2)^{2}}(\!s\!)
	&A^{*}_{(s-1)^{2}+1(s-2)^{2}}(\!s\!)&\cdots&A^{*}_{s^{2}(s-2)^{2}}(\!s\!)\\
	A_{\!(\!s\!-\!2\!)^{2}\!+1\!1\!}(\!s\!)&\cdots&A_{\!(\!s-2\!)^{2}+1(\!s-2\!)^{2}\!}(\!s\!)&A_{\!(\!s-2\!)^{2}+1(\!s-2\!)^{2}+1\!}(\!s\!)&\cdots
	&A^{*}_{\!(\!s-1\!)^{2}(\!s-2\!)^{2}+1\!}(\!s\!)&A^{*}_{\!(\!s-1\!)^{2}+1(\!s-2\!)^{2}+1\!}(\!s\!)&\cdots&A^{*}_{\!s^{2}(\!s-2\!)^{2}+1\!}(\!s\!)\\
	\cdots&\cdots&\cdots&\cdots&\cdots&\cdots&\cdots&\cdots&\cdots\\
	A_{\!(\!s\!-\!1\!)^{2}1}(\!s\!)&\cdots&A_{(s-1)^{2}(s-2)^{2}}(\!s\!)&A_{\!(\!s-1\!)^{2}(\!s-2\!)^{2}+1\!}(\!s\!)&\cdots&A_{(s-1)^{2}(s-1)^{2}}(\!s\!)
	&A^{*}_{(s-1)^{2}+1(s-1)^{2}}(\!s\!)&\cdots&A^{*}_{s^{2}(s-1)^{2}}(\!s\!)\\
	A_{\!(\!s\!-\!1\!)^{2}\!+1\!1\!}(\!s\!)&\cdots&A_{\!(\!s-1\!)^{2}+1(\!s-2\!)^{2}\!}(\!s\!)&A_{\!(\!s-1\!)^{2}+1(\!s-2\!)^{2}+1\!}(\!s\!)
	&\cdots&A_{\!(\!s-1\!)^{2}+1(\!s-1\!)^{2}\!}(\!s\!)&A_{\!(\!s-1\!)^{2}+1(\!s-1\!)^{2}+1\!}(\!s\!)&\cdots&A^{*}_{\!s^{2}(\!s-1\!)^{2}+1\!}(\!s\!)\\
	\cdots&\cdots&\cdots&\cdots&\cdots&\cdots&\cdots&\cdots&\cdots\\
	A_{s^{2}1}(s)&\cdots&A_{s^{2}(s-2)^{2}}(s)&A_{s^{2}(s-2)^{2}+1}(s)&\cdots&A_{s^{2}(s-1)^{2}}(s)&A_{s^{2}(s-1)^{2}+1}(s)&\cdots&A_{s^{2}s^{2}}(s)\\
	\end{array}
	\right]
	\end{equation}
\end{tiny}
where $A_{11}(s)$, $A_{22}(s)$, $\cdots$ $A_{s^{2}-1s^{2}-1}(s)$ and $A_{s^{2}s^{2}}(s)$ are real numbers. From (43)-(47), (73)-(75) and (77),
\begin{small}
	\begin{equation}
	\left \{\setlength\arraycolsep{0em}  \begin{array}{c}
	\frac{A_{\!s^{2}\!-\!1(\!s\!-\!2)^{2}\!+1\!}(\!s\!)}{\sqrt{1(2s-3)}}\!=\!\frac{A_{\!s^{2}\!-\!2(\!s\!-\!2\!)^{2}\!+2}(\!s\!)}{\sqrt{2(2s-4)}}
	\!=\!\cdots\!=\!\frac{A_{\!s^{2}\!-\!(\!s\!-\!1\!) (\!s\!-\!2\!)^{2}\!+s-1\!}(\!s\!)}{\sqrt{(s-1)(s-1)}}\!
	=\!\cdots\!=\!\frac{A_{\!s^{2}\!-\!(\!2s\!-\!3\!)(\!s\!-\!2\!)^{2}\!+\!2s\!-3}(\!s\!)}{\sqrt{(2s-3)1}} \\
	\frac{A_{\!(\!s\!-\!1\!)^{2}-1(\!s\!-\!3\!)^{2}+1}(\!s\!)}{\sqrt{1(2s-3)}}=\cdots
	=\frac{A_{\!(\!s\!-\!1\!)^{2}\!-\!(\!s\!-\!2\!)(\!s\!-\!3\!)^{2}\!+s-2}(\!s\!)}{\sqrt{(s-2)(s-2)}}=\cdots
	=\frac{A_{\!(\!s\!-\!1\!)^{2}\!-\!(\!2s\!-\!5\!)(\!s\!-\!3\!)^{2}\!+\!2s\!-\!5\!}(\!s\!)}{\sqrt{(2s-5)1}} \\
	\cdots, \cdots,\\
	\frac{A_{82}(s)}{\sqrt{1\cdot3}}=\frac{A_{73}(s)}{\sqrt{2\cdot2}}=\frac{A_{64}(s)}{\sqrt{3\cdot1}}
	\end{array}\right.
	\end{equation}
\end{small}
\begin{tiny}
	\begin{displaymath}\setlength\arraycolsep{0.1em}
	\begin{array}{cccccccccccccccc}\hat{A}_{z}[\psi_{\!s0\!0\!}&\psi_{\!s1\!-\!1\!}&\psi_{\!s10}&\cdots&\psi_{\!ss\!-\!33\!-\!s}&\psi_{\!ss\!-\!22\!-\!s}
	&\psi_{\!ss\!-\!23\!-\!s}&\cdots&\psi_{\!ss\!-\!2s\!-\!2}&\psi_{\!ss\!-\!1s\!-\!1}&\psi_{\!ss\!-\!1s\!-\!2}&\psi_{\!ss\!-\!1s\!-\!3}&\cdots
	&\psi_{\!ss\!-\!13\!-\!s}&\psi_{\!ss\!-\!12\!-\!s}&\psi_{\!ss\!-\!11\!-\!s}] \end{array}
	\end{displaymath}
\end{tiny}
\begin{tiny}
	\begin{displaymath}\setlength\arraycolsep{0.1em}
	=\begin{array}{cccccccccccccccc}[\psi_{\!s0\!0\!}&\psi_{\!s1\!-\!1\!}&\psi_{\!s10}&\cdots&\psi_{\!ss\!-\!33\!-\!s}&\psi_{\!ss\!-\!22\!-\!s}
	&\psi_{\!ss\!-\!23\!-\!s}&\cdots&\psi_{\!ss\!-\!2s\!-\!2}&\psi_{\!ss\!-\!1s\!-\!1}&\psi_{\!ss\!-\!1s\!-\!2}&\psi_{\!ss\!-\!1s\!-\!3}&\cdots
	&\psi_{\!ss\!-\!13\!-\!s}&\psi_{\!ss\!-\!12\!-\!s}&\psi_{\!ss\!-\!11\!-\!s}] \end{array}
	\end{displaymath}
\end{tiny}
\begin{tiny}
	\begin{equation}
	\left[\setlength\arraycolsep{0em}
	\begin{array}{cccccccccccccccc}
	0&0&T^{\!*}_{\!1}&\cdots&0&0&0&\cdots&0&0&0&0&\cdots&0&0&0\\
	0&0&0&\cdots&0&0&0&\cdots&0&0&0&0&\cdots&0&0&0\\
	T_{\!1}&0&0&\cdots&0&0&0&\cdots&0&0&0&0&\cdots&0&0&0\\
	\cdots&\cdots&\cdots&\cdots&\cdots&\cdots&\cdots&\cdots&\cdots&\cdots&\cdots&\cdots&\cdots&\cdots&\cdots&\cdots\\
	0&0&0&\cdots&0&0&\frac{\sqrt{\!(\!2\!s\!-\!5\!)1\!}}{s-2}T^{*}_{\!s\!-\!2\!}&\cdots&0&0&0&0&\cdots&0&0&0\\
	0&0&0&\cdots&0&0&0&\cdots&0&0&0&0&\cdots&0&\frac{\sqrt{\!1(\!2\!s\!-\!3\!)\!}}{s-1}T^{*}_{\!s\!-\!1\!}&0\\
	0&0&0&\cdots&\frac{\sqrt{\!(\!2\!s\!-\!5\!)1}}{s-2}T_{\!s\!-\!2\!}&0&0&\cdots&0&0&0&0&\cdots
	&\frac{\sqrt{\!2(\!2\!s\!-\!4\!)\!}}{s-1}T^{*}_{\!s\!-\!1\!}&0&0\\
	\cdots&\cdots&\cdots&\cdots&\cdots&\cdots&\cdots&\cdots&\cdots&\cdots&\cdots&\cdots&\cdots&\cdots&\cdots&\cdots\\
	0&0&0&\cdots&0&0&0&\cdots&0&0&\frac{\sqrt{\!(\!2\!s\!-\!3\!)1\!}}{s-1}T^{*}_{\!s\!-\!1\!}&0&\cdots&0&0&0\\
	0&0&0&\cdots&0&0&0&\cdots&0&0&0&0&\cdots&0&0&0\\
	0&0&0&\cdots&0&0&0&\cdots&\frac{\sqrt{\!(\!2\!s\!-\!3\!)1\!}}{s-1}T_{\!s\!-\!1\!}&0&0&0&\cdots&0&0&0\\
	0&0&0&\cdots&0&0&0&\cdots&0&0&0&0&\cdots&0&0&0\\
	\cdots&\cdots&\cdots&\cdots&\cdots&\cdots&\cdots&\cdots&\cdots&\cdots&\cdots&\cdots&\cdots&\cdots&\cdots&\cdots\\
	0&0&0&\cdots&0&0&\frac{\sqrt{\!2(\!2\!s\!-\!4\!)\!}}{s-1}T_{\!s\!-\!1\!}&\cdots&0&0&0&0&\cdots&0&0&0\\
	0&0&0&\cdots&0&\frac{\sqrt{\!1(\!2\!s\!-\!3\!)\!}}{s-1}T_{\!s\!-\!1\!}&0&\cdots&0&0&0&0&\cdots&0&0&0\\
	0&0&0&\cdots&0&0&0&\cdots&0&0&0&0&\cdots&0&0&0\\
	\end{array}
	\right]
	\end{equation}
\end{tiny}
\begin{tiny}
	\begin{displaymath}\setlength\arraycolsep{0.1em}
	\begin{array}{cccccccccccccccc}\hat{A}_{+}[\psi_{\!s0\!0\!}&\psi_{\!s1\!-\!1\!}&\psi_{\!s10}&\cdots&\psi_{\!ss\!-\!33\!-\!s}&\psi_{\!ss\!-\!22\!-\!s}
	&\psi_{\!ss\!-\!23\!-\!s}&\cdots&\psi_{\!ss\!-\!2s\!-\!2}&\psi_{\!ss\!-\!1s\!-\!1}&\psi_{\!ss\!-\!1s\!-\!2}&\psi_{\!ss\!-\!1s\!-\!3}&\cdots
	&\psi_{\!ss\!-\!13\!-\!s}&\psi_{\!ss\!-\!12\!-\!s}&\psi_{\!ss\!-\!11\!-\!s}] \end{array}
	\end{displaymath}
\end{tiny}
\begin{tiny}
	\begin{displaymath}\setlength\arraycolsep{0.1em}
	=\begin{array}{cccccccccccccccc}[\psi_{\!s0\!0\!}&\psi_{\!s1\!-\!1\!}&\psi_{\!s10}&\cdots&\psi_{\!ss\!-\!33\!-\!s}&\psi_{\!ss\!-\!22\!-\!s}
	&\psi_{\!ss\!-\!23\!-\!s}&\cdots&\psi_{\!ss\!-\!2s\!-\!2}&\psi_{\!ss\!-\!1s\!-\!1}&\psi_{\!ss\!-\!1s\!-\!2}&\psi_{\!ss\!-\!1s\!-\!3}&\cdots
	&\psi_{\!ss\!-\!13\!-\!s}&\psi_{\!ss\!-\!12\!-\!s}&\psi_{\!ss\!-\!11\!-\!s}] \end{array}
	\end{displaymath}
\end{tiny}
\begin{tiny}
	\begin{equation}
	\left[\setlength\arraycolsep{0em}
	\begin{array}{cccccccccccccccc}
	0&\sqrt{\!2}T^{\!*}_{\!1}&0&\cdots&0&0&0&\cdots&0&0&0&0&\cdots&0&0&0\\
	0&0&0&\cdots&0&0&0&\cdots&0&0&0&0&\cdots&0&0&0\\
	0&0&0&\cdots&0&0&0&\cdots&0&0&0&0&\cdots&0&0&0\\
	\cdots&\cdots&\cdots&\cdots&\cdots&\cdots&\cdots&\cdots&\cdots&\cdots&\cdots&\cdots&\cdots&\cdots&\cdots&\cdots\\
	0&0&0&\cdots&0&\frac{\sqrt{\!(\!2\!s\!-\!5\!)(\!2\!s\!-\!4\!)\!}}{s-2}T^{\!*}_{\!s\!-\!2\!}&0&\cdots&0&0&0&0&\cdots&0&0&0\\
	0&0&0&\cdots&0&0&0&\cdots&0&0&0&0&\cdots&0&0&\frac{\sqrt{\!(\!2\!s\!-\!3\!)(\!2\!s\!-\!2\!)\!}}{s-1}T^{\!*}_{\!s\!-\!1\!}\\
	0&0&0&\cdots&0&0&0&\cdots&0&0&0&0&\cdots&0&\frac{\sqrt{\!(\!2\!s\!-\!4\!)(\!2\!s\!-\!3\!)\!}}{s-1}T^{\!*}_{\!s\!-\!1\!}&0\\
	\cdots&\cdots&\cdots&\cdots&\cdots&\cdots&\cdots&\cdots&\cdots&\cdots&\cdots&\cdots&\cdots&\cdots&\cdots&\cdots\\
	0&0&0&\cdots&0&0&0&\cdots&0&0&0&\frac{\sqrt{2}}{\!s\!-\!1\!}T^{\!*}_{\!s\!-\!1\!}&\cdots&0&0&0\\
	0&0&0&\cdots&0&0&0&\cdots&-\!\frac{\sqrt{\!(\!2\!s\!-\!3\!)(\!2\!s\!-\!2\!)}}{s-1}T_{\!s\!-\!1\!}&0&0&0&\cdots&0&0&0\\
	0&0&0&\cdots&0&0&0&\cdots&0&0&0&0&\cdots&0&0&0\\
	0&0&0&\cdots&0&0&0&\cdots&0&0&0&0&\cdots&0&0&0\\
	\cdots&\cdots&\cdots&\cdots&\cdots&\cdots&\cdots&\cdots&\cdots&\cdots&\cdots&\cdots&\cdots&\cdots&\cdots&\cdots\\
	0&0&0&\cdots&0&-\frac{\sqrt{2}}{s-1}T_{s-1}&0&\cdots&0&0&0&0&\cdots&0&0&0\\
	0&0&0&\cdots&0&0&0&\cdots&0&0&0&0&\cdots&0&0&0\\
	0&0&0&\cdots&0&0&0&\cdots&0&0&0&0&\cdots&0&0&0\\
	\end{array}
	\right]
	\end{equation}
\end{tiny}
\begin{tiny}
	\begin{displaymath}\setlength\arraycolsep{0.1em}
	\begin{array}{cccccccccccccccc}\hat{A}_{-}[\psi_{\!s0\!0\!}&\psi_{\!s1\!-\!1\!}&\psi_{\!s10}&\cdots&\psi_{\!ss\!-\!33\!-\!s}&\psi_{\!ss\!-\!22\!-\!s}
	&\psi_{\!ss\!-\!23\!-\!s}&\cdots&\psi_{\!ss\!-\!2s\!-\!2}&\psi_{\!ss\!-\!1s\!-\!1}&\psi_{\!ss\!-\!1s\!-\!2}&\psi_{\!ss\!-\!1s\!-\!3}&\cdots
	&\psi_{\!ss\!-\!13\!-\!s}&\psi_{\!ss\!-\!12\!-\!s}&\psi_{\!ss\!-\!11\!-\!s}] \end{array}
	\end{displaymath}
\end{tiny}
\begin{tiny}
	\begin{displaymath}\setlength\arraycolsep{0.1em}
	=\begin{array}{cccccccccccccccc}[\psi_{\!s0\!0\!}&\psi_{\!s1\!-\!1\!}&\psi_{\!s10}&\cdots&\psi_{\!ss\!-\!33\!-\!s}&\psi_{\!ss\!-\!22\!-\!s}
	&\psi_{\!ss\!-\!23\!-\!s}&\cdots&\psi_{\!ss\!-\!2s\!-\!2}&\psi_{\!ss\!-\!1s\!-\!1}&\psi_{\!ss\!-\!1s\!-\!2}&\psi_{\!ss\!-\!1s\!-\!3}&\cdots
	&\psi_{\!ss\!-\!13\!-\!s}&\psi_{\!ss\!-\!12\!-\!s}&\psi_{\!ss\!-\!11\!-\!s}] \end{array}
	\end{displaymath}
\end{tiny}
\begin{tiny}
	\begin{equation}
	\left[\setlength\arraycolsep{0em}
	\begin{array}{cccccccccccccccc}
	0&0&0&\cdots&0&0&0&\cdots&0&0&0&0&\cdots&0&0&0\\
	\sqrt{\!2}T_{\!1}&0&0&\cdots&0&0&0&\cdots&0&0&0&0&\cdots&0&0&0\\
	0&0&0&\cdots&0&0&0&\cdots&0&0&0&0&\cdots&0&0&0\\
	\cdots&\cdots&\cdots&\cdots&\cdots&\cdots&\cdots&\cdots&\cdots&\cdots&\cdots&\cdots&\cdots&\cdots&\cdots&\cdots\\
	0&0&0&\cdots&0&0&0&\cdots&0&0&0&0&\cdots&0&0&0\\
	0&0&0&\cdots&\frac{\sqrt{\!(2\!s\!-\!5\!)(\!2\!s\!-\!4\!)}}{s-2}T_{\!s\!-\!2\!}&0&0&\cdots&0&0&0&0
	&\cdots&\frac{\!-\sqrt{\!2\!}}{s-1}T^{*}_{\!s\!-\!1\!}&0&0\\
	0&0&0&\cdots&0&0&0&\cdots&0&0&0&0&\cdots&0&0&0\\
	\cdots&\cdots&\cdots&\cdots&\cdots&\cdots&\cdots&\cdots&\cdots&\cdots&\cdots&\cdots&\cdots&\cdots&\cdots&\cdots\\
	0&0&0&\cdots&0&0&0&\cdots&0&\!-\!\frac{\sqrt{\!(\!2\!s\!-\!3\!)(\!2\!s\!-\!2\!)}}{s-1}T^{*}_{\!s\!-\!1\!}&0&0&\cdots&0&0&0\\
	0&0&0&\cdots&0&0&0&\cdots&0&0&0&0&\cdots&0&0&0\\
	0&0&0&\cdots&0&0&0&\cdots&0&0&0&0&\cdots&0&0&0\\
	0&0&0&\cdots&0&0&0&\cdots&\frac{\sqrt{2}}{\!s\!-\!1\!}T_{\!s\!-\!1\!}&0&0&0&\cdots&0&0&0\\
	\cdots&\cdots&\cdots&\cdots&\cdots&\cdots&\cdots&\cdots&\cdots&\cdots&\cdots&\cdots&\cdots&\cdots&\cdots&\cdots\\
	0&0&0&\cdots&0&0&0&\cdots&0&0&0&0&\cdots&0&0&0\\
	0&0&0&\cdots&0&0&\frac{\sqrt{\!(\!2\!s\!-\!4\!)(\!2\!s\!-\!3\!)\!}}{s-1}T_{\!s\!-\!1\!}&\cdots&0&0&0&0&\cdots&0&0&0\\
	0&0&0&\cdots&0&\frac{\sqrt{\!(\!2\!s\!-\!3\!)(\!2\!s\!-\!2\!)}}{s-1}T_{\!s\!-\!1\!}&0&\cdots&0&0&0&0&\cdots&0&0&0\\
	\end{array}
	\right]
	\end{equation}
\end{tiny}
with $T_{s-1}=A_{s^{2}-(s-1)(s-2)^{2}+s-1}(s)$, $T_{s-2}=A_{(s-1)^{2}-(s-2)(s-3)^{2}+s-2}(s)$, $\cdots$, $T_{1}=A_{31}(s)$. From (48), (73), (76) and (79)-(81),
\begin{displaymath}
(2\hbar^{2}s^2E_{s}+\mu k^{2})\psi_{ss-1s-1}=0
\end{displaymath}
So
\begin{equation}
E_{s}=-\frac{\mu k^2}{2\hbar^2}\frac{1}{s^2}
\end{equation}
From (49), (73) and (80)-(81),
\begin{displaymath}
|A_{s^{2}-(s-1)(s-2)^{2}+(s-1)}(s)|^{2}=\frac{(s-1)^{2}}{2s-3}\frac{k^{2}}{s^{2}}
\end{displaymath}
\begin{footnotesize}
	\begin{displaymath}
	|A_{(s-1)^{2}-(s-2)(s-3)^{2}+(s-2)}(s)|^{2}-\frac{(2s-1)(s-2)^{2}}{(2s-5)(s-1)^{2}}|A_{s^{2}-(s-1)(s-2)^{2}+(s-1)}(s)|^{2}
	=\frac{(s-2)^{2}}{2s-5}\frac{k^{2}}{s^{2}}
	\end{displaymath}
\end{footnotesize}
\begin{displaymath}
\cdots, \cdots,
\end{displaymath}
\begin{displaymath}
|A_{31}(s)|^{2}-\frac{5}{4}|A_{73}(s)|^{2}=\frac{k^{2}}{s^{2}}
\end{displaymath}
Thus
\begin{small}
	\begin{equation}
	\left \{\!
	\begin{array}{c}
	|A_{s^{2}-(s-1)(s-2)^{2}+(s-1)}(s)|^{2}=\frac{2s-1}{(2s-1)(2s-3)}(s\!-\!1)^{2}\frac{k^{2}}{s^{2}}=\frac{s^{2}-(s-1)^{2}}{4(s-1)^{2}-1}(s\!-\!1)^{2}
	\frac{k^{2}}{s^{2}}\\
	\!|A_{\!(\!s-\!1)^{2}\!-\!(\!s-\!2)(\!s-\!3)^{2}+(\!s-\!2)}(s)|^{2}\!=\!\frac{(2s-1)+(2s-3)}{(2s-3)(2s-5)}(\!s\!-\!2)^{2}\frac{k^{2}}{s^{2}}
	\!=\!\frac{s^{2}-(s-2)^{2}}{4(s-2)^{2}-1}(\!s\!-\!2)^{2}\frac{k^{2}}{s^{2}} \\
	\cdots,\cdots,\\
	|A_{31}(s)|^{2}=\frac{(2s-1)+(2s-1)+\cdots+3}{3\cdot1}\frac{k^{2}}{s^{2}}=\frac{s^{2}-1}{4-1}\frac{k^{2}}{s^{2}}
	\end{array}
	\right.
	\end{equation}
\end{small}
If we take positive real solutions from (83), then (79)-(81) become
\begin{equation}
\left \{  \begin{array}{c}
\hat{A}_{z}\psi_{s00}=\frac{k}{s}\sqrt{\frac{s^{2}-1}{4-1}}\psi_{s10} \\
\cdots,\cdots,\\
\hat{A}_{z}\psi_{ss\!-\!1s\!-\!1}=0,\cdots,\hat{A}_{z}\psi_{ss-11-s}=0
\end{array}\right.
\end{equation}
\begin{equation}
\left \{  \begin{array}{c}
\hat{A}_{+}\psi_{s00}=-\frac{k}{s}\sqrt{\frac{s^{2}-1}{4-1}}\psi_{s11} \\
\cdots,\cdots,\\
\hat{A}_{\!+}\psi_{ss\!-\!1s\!-\!1}\!=\!0,\cdots,
\hat{A}_{\!+}\psi_{ss\!-\!11\!-\!s}\!=\!\frac{k}{s}\sqrt{\frac{s^{2}-(s-1)^{2}}{4(s-1)^{2}-1}(2s\!-\!2)(2s\!-\!3)}\psi_{ss\!-\!22\!-\!s}
\end{array}\right.
\end{equation}
\begin{equation}
\left \{  \begin{array}{c}
\hat{A}_{-}\psi_{s00}=\frac{k}{s}\sqrt{\frac{s^{2}-1}{4-1}2}\psi_{s1-1} \\
\cdots,\cdots,\\
\hat{A}_{-}\psi_{ss\!-\!1s\!-\!1}\!=\!\frac{k}{s}\sqrt{\frac{s^{2}-(s-1)^{2}}{4(s-1)^{2}-1}(2s\!-\!2)(2s\!-\!3)}\psi_{ss\!-\!2s\!-\!2},
\cdots,\hat{A}_{\!-}\psi_{\!ss\!-\!11\!-\!s}\!=\!0
\end{array}\right.
\end{equation}
Therefore
\begin{equation}
E_{n}=-\frac{\mu k^{2}}{2\hbar^{2}}\frac{1}{n^{2}}
\end{equation}
\begin{footnotesize}
	\begin{equation}
	\hat{A}_{z}\psi_{nlm}=\frac{k}{n}\sqrt{\frac{n^{2}-l^{2}}{4l^{2}-1}(l^{2}-m^{2})}\psi_{nl-1m}
	+\frac{k}{n}\sqrt{\frac{n^{2}-(l+1)^{2}}{4(l+1)^{2}-1}[(l+1)^{2}-m^{2}]}\psi_{nl+1m}
	\end{equation}
\end{footnotesize}
\begin{scriptsize}
	\begin{equation}
	\hat{A}_{\!+}\psi_{\!nlm}\!=\!\frac{k}{n}\!\sqrt{\frac{n^{2}\!-\!l^{2}}{\!4l^{2}\!-\!1\!}(l\!-\!m\!)(l\!-\!m\!-\!1\!)}\psi_{\!nl\!-\!1m\!+1}
	\!-\!\frac{k}{n}\!\sqrt{\frac{n^{2}\!-\!(l\!+\!1\!)^{2}}{\!4(l\!+\!1\!)^{2}\!-\!1\!}(l\!+\!m\!+\!1)(l\!+\!m\!+\!2)}\psi_{\!nl\!+\!1m\!+\!1}
	\end{equation}
\end{scriptsize}
\begin{scriptsize}
	\begin{equation}
	\hat{A}_{\!-}\psi_{\!nlm}\!=\!\frac{k}{n}\!\sqrt{\frac{n^{2}\!-\!(l\!+\!1)^{2}}{4(l\!+\!1)^{2}\!-\!1}(l\!-\!m\!+\!1)(l\!-\!m\!+\!2)}\psi_{\!nl\!+\!1m\!-\!1}
	\!-\!\frac{k}{n}\!\sqrt{\!\frac{n^{2}\!-\!l^{2}}{4l^{2}\!-\!1}(l\!+\!m)(l\!+\!m\!-\!1)}\psi_{\!nl\!-\!1m\!-\!1}
	\end{equation}
\end{scriptsize}
with
\begin{displaymath}
n=1,2,3,\dots;l=0,1,\dots,n-1;m=-l,1-l,\dots,l
\end{displaymath}
When $l=n-1$ and $m=-l$,
\begin{displaymath}
\begin{array}{cc}
\hat{A}_{z}(R_{nn-1}Y_{n-11-n})=0 \Rightarrow \frac{dR_{nn-1}}{dr}=(\frac{n-1}{r}-\frac{1}{na})R_{nn-1} & (a=\frac{\hbar^{2}}{\mu k})
\end{array}
\end{displaymath}
Solving this equation, we will find $R_{nn-1}$.
\begin{equation}
\begin{array}{cc}
R_{nn-1}=\frac{1}{\sqrt{(2n)!}}(\frac{2}{na})^{n+\frac{1}{2}}r^{n-1}e^{-\frac{r}{na}} & (n=1,2,\cdots)
\end{array}
\end{equation}
When $m=1-l$, from (88),
\begin{displaymath}
\sqrt{\frac{n^{2}-l^{2}}{2l+1}}R_{nl-1}Y_{l-11-l}=\frac{n}{k}\hat{A}_{z}(R_{nl}Y_{l1-l})-\sqrt{\frac{n^{2}-(l+1)^{2}}{4(l+1)^{2}-1}4l}R_{nl+1}Y_{l+11-l}
\end{displaymath}
Furthermore,
\begin{displaymath}
\begin{array}{cc}
R_{nl-1}=\frac{(2l+1)na}{\sqrt{n^{2}-l^{2}}}\frac{dR_{nl}}{dr}+\sqrt{\frac{n^{2}-(l+1)^{2}}{n^{2}-l^{2}}}R_{nl+1} & (l=n-1,n-2,\cdots,1)
\end{array}
\end{displaymath}
and
\begin{displaymath}
\begin{array}{cc}
R_{nl+1}=\frac{n(l+1)}{\sqrt{n^{2}-(l+1)^{2}}}[(\frac{a}{r}l-\frac{1}{l+1})R_{nl}-a\frac{d R_{nl}}{dr}] & (l=0,1,\cdots,n-2)
\end{array}
\end{displaymath}
Thus
\begin{equation}
\begin{array}{cc}
R_{nl-1}=\frac{nla}{\sqrt{n^{2}-l^{2}}}[\frac{dR_{nl}}{dr}+(\frac{l+1}{r}-\frac{1}{al})R_{nl}] & (l=n-1,n-2,\cdots,1)
\end{array}
\end{equation}
We can get $R_{10},R_{21},R_{32},\cdots$ from (91) and other radial wave functions are obtained from (92).\\
When $n=1$,
\begin{displaymath}
R_{10}=\frac{2}{a^{\frac{3}{2}}}e^{-\frac{r}{a}}
\end{displaymath}
When $n=2$,
\begin{displaymath}
R_{21}=\frac{1}{2\sqrt{6}a^{\frac{5}{2}}}r e^{-\frac{r}{2a}}; R_{20}=\frac{1}{\sqrt{2}a^{\frac{3}{2}}}(1-\frac{r}{2a})e^{-\frac{r}{2a}}
\end{displaymath}
When $n=3$,
\begin{displaymath}
R_{32}=\frac{4}{81\sqrt{30}a^{\frac{7}{2}}}r^{2}e^{-\frac{r}{3a}};
\end{displaymath}
\begin{displaymath}
R_{31}=\frac{4}{27\sqrt{6}a^{\frac{5}{2}}}(2-\frac{r}{3a})re^{-\frac{r}{3a}},
R_{30}=\frac{2}{3\sqrt{3}a^{\frac{3}{2}}}(1-\frac{2r}{3a}+\frac{2r^{2}}{27a^{2}})e^{-\frac{r}{3a}}
\end{displaymath}
$\cdots, \cdots$,\\
Let $y=\frac{2r}{na}$, by means of mathematical induction, the following expression is proved from (91) and (92).
\begin{equation}
\begin{array}{cc}
R_{nl}=\sqrt{(\frac{2}{na})^{3}\frac{(n-1-l)!}{2n(n+l)!}}y^{l}L_{n-1-l}^{2l+1}(y)e^{-\frac{y}{2}} & (l=n\!-\!1,\cdots,1,0)
\end{array}
\end{equation}
where $L_{n-1-l}^{2l+1}(y)$ are the associated Laguerre polynomial and $L_{n-1-l}^{2l+1}(y)=\frac{y^{-2l-1}}{(n-1-l)!}e^{y}\frac{d^{n-1-l}}{dy^{n-1-l}}(e^{-y}y^{n+l})$.
Thus the previous work of Heisenberg et al on matrix mechanics and of Schr$\ddot{o}$dinger on wave mechanics will be incorporated into a single mathematical formalism. As a result, the descriptions of matrix mechanics and wave mechanics on one-dimensional harmonic oscillator and the hydrogen atom have been unified here. These methods and conclusions can be generalized to the whole of quantum mechanics.

\textbf{Author contribution statement}
Yongqin Wang drafted the manuscript. Lifeng Kang revised the manuscript and edited the English language.\\
\end{document}